\documentclass[prl,amsmath,amssymb,showpacs,groupaddress,twocolumn]{revtex4}
\usepackage{graphicx}
\usepackage{bm}

\newcommand*{\J}{{\cal J}}
\newcommand*{\x}{{\text{x}}}
\newcommand*{\E}{{\cal E}}
\newcommand*{\F}{{\cal F}}
\newcommand*{\K}{{\cal K}}
\newcommand*{\ud}{{\uparrow\downarrow}}
\renewcommand*{\u}{{\uparrow}}
\renewcommand*{\d}{{\downarrow}}
\renewcommand*{\ni} { \text{Ni}}
\newcommand*{\cu} { \text{Cu}}
\newcommand{\m} {\mathfrak{m}}
\begin{document}

\title{Voltage-Controlled Surface Magnetization of Itinerant Ferromagnet Ni$_\text{1-x}$Cu$_\text{x}$}
\author{Igor V. Ovchinnikov}
\email{iovchinnikov@ucla.edu}
\author{Kang L. Wang}
\email{wang@ee.ucla.edu} \affiliation{Electrical Engineering Department, University of California at Los Angeles, Los Angeles, CA, 90095-1594}
\begin{abstract}
We argue that surface magnetization of a metallic ferromagnet can be turned on and off isothermally by an applied voltage. For this, the material's electron subsystem must be close enough to the boundary between para- and ferromagnetic regions on the electron density scale. For the 3d series, the boundary is between Ni and Cu, which makes their alloy a primary candidate. Using Ginzburg-Landau functional, which we build from Ni$_\text{1-x}$Cu$_\x$ empirical properties, ab-initio parameters of Ni and Cu, and orbital-free LSDA, we show that the proposed effect is experimentally observable.
\end{abstract}
 \maketitle

Further evolution of magneto-electronics \cite{SpintronicsReview} depends highly on the availability of materials, in which local magnetization can be turned on and off isothermally by an electric voltage.
The hopes to achieve this are mainly laid on the dilute magnetic semiconductors (DMS) \cite{DMSGeneral},
in which the effect was recently demonstrated experimentally \cite{DMSDiscovery}. The voltage-controlled
ferromagnetic ordering in DMS relies on the virtue of doped semiconductors to allow external variation
of the free-carrier spatial density within the semiconductor depletion layer, typically measured in
dozens of nanometers. The voltage variation of the high electron density in metals is possible only
within the atomic size Thomas-Fermi (TF) surface layer. As a result, the voltage controlled
ferromagnetism in a metal has not been considered a possibility lately.

We argue that by capacitively charging a metallic ferromagnet one can drive the surface electron
subsystem in and out of its ferromagnetic state. At this, the electron system of the metal has to be
paramagnetic at the device operation temperature (room temperature, $T_0$), but close enough to the
ferromagnetic state on the temperature and/or electron density scales. The proximity of the
ferromagnetic transition will play a twofold role: (\emph{i}) the capacitive change in the electron
density is relatively small, so that the transition has to be sufficiently close in order to reach the
ferromagnetic region with reasonable voltages, and (\emph{ii}) due to the critical collective spin
correlations, the spin correlation length grows infinitely as one approaches the transition point.
Consequently, even though the injected carriers are spatially limited to the TF-layer, the system must
develop a much wider surface magnetization profile. In this paper we investigate the proposed
possibility in $\ni_\text{1-x}\cu_\text{x}$.

\begin{figure}[t]
\includegraphics[width=8.6 cm,height=6cm]{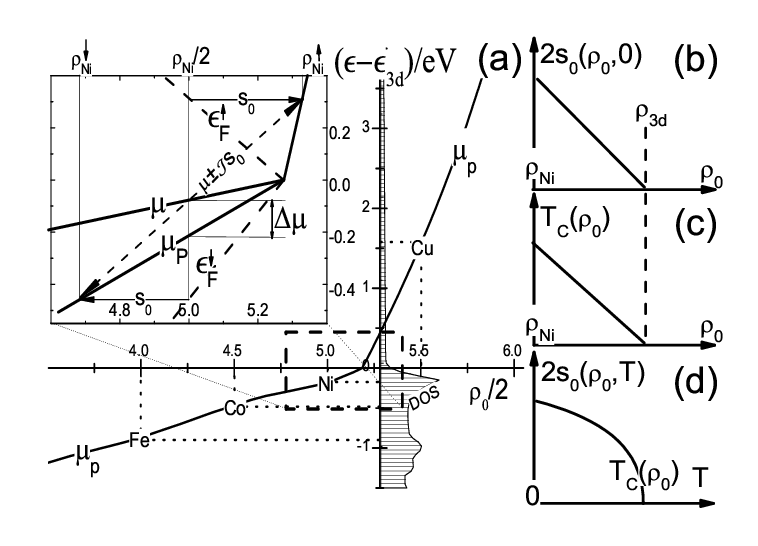}
\caption{\label{Figure1} (a) Schematics showing the paramagnetic chemical
potentials of the elemental ferromagnets on the kinetic energy, $\epsilon$,
and the spatial electron density, $\rho_0$ (outer shell electrons per atom),
scales. The inset is the magnified dashed area, which shows the graphic
solution for Eqs.(\ref{equations1}) determining the spin density, $s_0$, the
paramagnetic and ferromagnetic chemical potentials, $\mu_\text{P}$ and $\mu$,
and the spin up and down \emph{kinetic} Fermi energies, $\varepsilon_F^\ud$,
as functions of $\rho_0$ for pure Ni, $\rho_0=\rho_\ni$. (b)-(d) The
experimentally observed properties of $\ni_\text{1-x}\cu_\x$: (b)
zero-temperature magnetization \emph{vs}. $\rho_0$; (c) Curie temperature
\emph{vs}. $\rho_0$; (d) the magnetization \emph{vs}. temperature, $T$.}
\end{figure}

The nature of ferromagnetism lies in the competition between the kinetic energy and the exchange
interaction. The kinetic energy of the spatial quantization tends to equalize the numbers of spin up and
down electrons by shifting the fermionic antisymmetry into the spin sector of the many-body
wavefunction. In turn, the exchange interaction does the opposite, struggling to unbalance the up and
down spins. For itinerant ferromagnets, - the subclass of materials elemental ferromagnets belong to, -
the outcome of this competition can be predicted from the Stoner criterion \cite{citeStoner}. According
to the Stoner criterion, ferromagnets possess a high density of states (DOS) at the (paramagnetic)
chemical potential. In the 3d series, the high DOS is provided by the 3d band on the background of the
low and wide 4s and 4p bands (see Fig.\ref{Figure1}a). Thus, the elements with the chemical potential
within the 3d band, Fe, Co, and Ni, are ferromagnetic, whereas the very next element, Cu, is not. Cu has
an extra electron per atom beyond Ni, so that on the electron density scale the ferro-to-para boundary
lies between Ni and Cu.

Despite their different magnetic properties, Ni and Cu are very similar from
the band structure point of view. In a crystalline state both form the
face-centered cubic lattice with almost the same lattice constant, $a$ ($6.69$
\emph{vs}. $6.83$ a.u.). In result, on alloying, Ni and Cu form substitutional
solid solutions at all compositions. Consequently,
Ni$_\text{1-x}$Cu$_\text{x}$ can be though of as an all-the-same structure
solid, with the equilibrium electron density varying linearly with x, $\rho_0
= \rho_\ni+\x/a^3$. Within this picture, $\x$ and $\rho_0$ are
interchangeable.

Other experimentally observed properties of the alloy \cite{NiCuAlloy} can be well approximated as: (see
Fig.\ref{Figure1}b-d) the linear dependence of the zero temperature spin density, $s_0(\rho_0,0)$, and
the Curie temperature, $T_C(\rho_0)$, on $\rho_0$, and the Landau dependence of the spin density,
$s_0(\rho_0,T)$, on the temperature, $T$:
\begin{subequations} \label{requirements}
\begin{eqnarray}
2s_0(\rho_0,0)/\gamma &=& (\rho_\text{3d}-\rho_0) \theta(\rho_\text{3d}-\rho_0),\label{linearmagnetization}\\
T_{C}(\rho_0)/\kappa &=& (\rho_\text{3d}-\rho_0) \theta(\rho_\text{3d}-\rho_0),\label{linearCurrierTemperature}\\
s_0(\rho_0,T)/s_0(\rho_0,0) &=& (1-\tilde T)^{1/2}\theta(\rho_\text{3d}-\rho_0),\label{LandauDependence}
\end{eqnarray}
\end{subequations}
where $\tilde T = T/T_{C}(\rho_0)$, $\theta$ is the Heaviside step-function,
$\rho_\text{3d}=\rho_\ni+0.53/a^3$ is the position of the $3d$ band edge on
the density scale, and $\gamma$ and $\kappa$ are the slopes of the
$2s_0(\rho_0,0)$ and $T_C(\rho_0)$ lines. For nickel
$2s_0(\rho_\ni,0)=0.66/a^3$ and $T_C(\rho_\ni)=627\text{K}$, so that
$\gamma\approx 1.25$ and $\kappa\approx1.2\times10^2\text{K}a^3$.

A rather accurate quantitative look at the itinerant ferromagnetism can be
taken via the Stoner approximation \cite{citeStoner}. In a homogeneous system
case, the zero-temperature energy per volume is the sum of the kinetic and
exchange parts:
\begin{eqnarray}
\E_{\cal KX}(\rho_0,s_0) = \sum\nolimits_\ud\K (\rho_0^\ud) - \J s_0^2,\label{zerotemparatureenergy}
\end{eqnarray}
where $\rho^\ud_0 = \rho_0 /2 \pm s_0$ are the up and down spin densities, $\J$ is the Heisenberg
exchange interaction constant \cite{MinHam}, and the kinetic energy is defined as $\K(\rho^\ud_0) =
\int_{-\infty}^{\epsilon_F^\ud} \epsilon' \nu(\epsilon') d\epsilon'$, with
$\epsilon_F^\ud\equiv\epsilon_F(\rho^\ud_0)$ being the \emph{kinetic} spin up and down Fermi energies,
and $\nu$ being the DOS (per spin). \footnote{To avoid confusion, even though
$\epsilon_F^\u>\epsilon_F^\d$ for ferromagnetic states, the total \emph{kinetic-exchange} Fermi energies
are, of course, the same and equal the chemical potential.} The function $\epsilon_F$ defines the
one-to-one correspondence between the (spin) density and the position of the Fermi energy on the kinetic
energy scale. It can be defined via its inverse function as $\rho =
\int_{-\infty}^{\epsilon_F}\nu(\epsilon') d\epsilon'$, so that $\partial \epsilon_F/\partial \rho =
\nu(\epsilon_F)^{-1}$ and $\partial\K/\partial \rho=\epsilon_F$, \emph{i.e.}, $\epsilon_F$ and $\rho$
are the Legandre conjugates with respect to $\K$.

For the ground state, vanishing variations of $\E_{\cal KX}$ in $\rho_0$ and $s_0$ yield the equations
determining the ground state's spin density and the chemical potential as functions of $\rho_0$:
\begin{eqnarray}
2 s_0(\rho_0) \J =  \epsilon_F^\u - \epsilon_F^\d, \text{ }
\mu(\rho_0)=(\epsilon_F^\u + \epsilon_F^\d )/2.\label{equations1}
\end{eqnarray}
To get (\ref{linearmagnetization}) from Eqs.(\ref{equations1}) (the graphic solution for pure Ni is
given in the inset of Fig.\ref{Figure1}a) the DOS has to be of the following simple form
($\epsilon_\text{3d}$ is the 3d band edge):
\begin{eqnarray*}
\nu (\epsilon)= \nu_\ni\theta(\epsilon_\text{3d}-\epsilon) + \nu_\cu\theta(\epsilon-\epsilon_\text{3d}).
\end{eqnarray*}
Then, $\gamma =(\nu_\cu^{-1}-\nu_\ni^{-1})/(\nu_\ni^{-1} + \nu_\cu^{-1} - 2\J)$, and
\begin{eqnarray}
\mu (\rho_0) = \mu_\text{P}(\rho_0) - \Delta \mu (\rho_0), \label{ZeroTemperatureChemicalPot}
\end{eqnarray}
where
\begin{eqnarray}
\mu_\text{P}(\rho_0) = \epsilon_\text{3d} + \frac12(\rho_0-\rho_\text{3d}) \times \left\{\begin{array}{cc}\nu^{-1}_\ni, & \rho_0<\rho_\text{3d},\\
\nu^{-1}_\cu, & \rho_0>\rho_\text{3d},\end{array}\right.\label{paramagneticChemicalPotential}
\end{eqnarray}
is the chemical potential, corresponding to the paramagnetic solution
($s_0=0,\mu_\text{P}=\mu=\epsilon^\ud_F=\epsilon_F(\rho_0/2)$), and
\begin{eqnarray}
\Delta\mu (\rho_0) = (2\Delta\nu)^{-1}(\rho_0-\rho_\text{3d})\theta(\rho_\text{3d}-\rho_0),
\label{ZeroTemperatureChemicalPotShift}
\end{eqnarray}
is the chemical potential shift due to the switching from the paramagnetic to
the ferromagnetic state, with $\Delta\nu^{-1}=\gamma (\J-\nu_\ni^{-1})$.
\footnote{In fact, the paramagnetic solution always exists. In the
ferromagnetic region, however, the paramagnetic solution is unstable with
respect to fluctuations in $s_0$ (the Stoner criterion): $\left.\partial^2
\E_{\cal KX} /\partial s_0^2\right|_{s_0=0} <0$.} Numerical estimates for the
parameters of the model can be obtained by fitting
$\mu_{\text{P}}(\rho_{\ni,\cu})-\epsilon_\text{3d}$ from
Eq.(\ref{paramagneticChemicalPotential}) and the exchange spitting in Ni, $2
s_0(\rho_\ni) \J$, with their ab-initio values
\cite{The_Book_On_Band_Structure}: $\nu^{-1}_{\cu,\ni} = 6.7, 0.79
\text{eV}a^3,\J = 1.3 \text{eV}a^3$, and $\Delta\nu^{-1} = 0.61\text{eV}a^3$.

\begin{figure}[t]
\includegraphics[width=8.6cm, height=6.6cm]{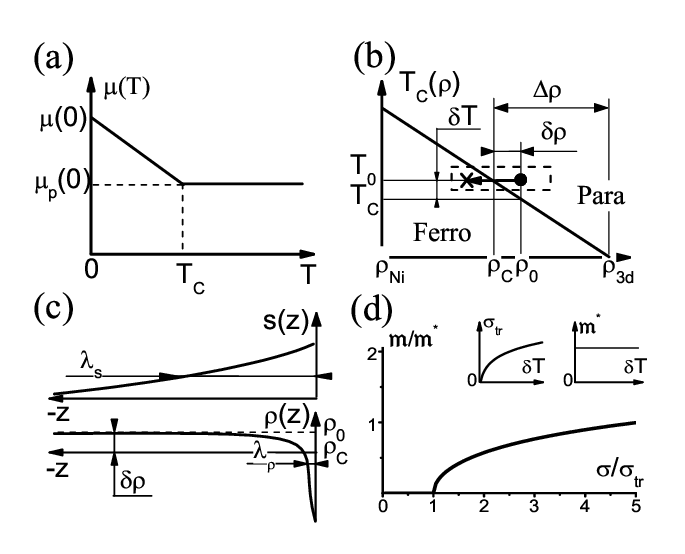}
\caption{\label{Figure2} (a) The chemical potential provided by the proposed model \emph{vs}.
temperature (b) The area on the ($T,\rho$)-plane under consideration (dashed rectangle). The parameters
$\Delta\rho$, $\rho_C$,$\delta\rho$,  and $\delta T$ are introduced in the text. The arrow symbolizes
the effect of the (positive) bias on the surface electron subsystem. (c) The widths of the spatial
profiles of the magnetization and the density are determined by the spin and density correlation lengths
(Eqs.\ref{lambda_s}), and $\lambda_s\gg\lambda_\rho$. (d) The surface magnetization, $\m$, \emph{vs.}
surface charge, $\sigma$. (insets) The threshold charging, $\sigma_\text{tr}$, and the charachteristic
magnetization, $\m^*$, \emph{vs.} the alloy's detuning from the transition, $\delta T$.}
\end{figure}

For non-zero temperatures, the free energy of a homogeneous system can be assumed a function of
$\rho_0,s_0$, and $T$. Its spatial density can be given as the sum of the paramagnetic and ferromagnetic
(Landau) parts:
\begin{subequations}
\label{FreeEnergy}
\begin{eqnarray}
\F_{\cal KX}(\rho_0,s_0,T) = \F_\text{P}(\rho_0,T) + \Delta\F (\rho_0,s_0,T).
\label{FreeEnergyGeneralForm}
\end{eqnarray}
$\F_\text{P}$ is determined mostly by the crystal structure. For not very high
temperatures, such that all the structural phase transitions are far on the
$(T,\rho)$-plane, $\F_\text{p}$ can be assumed temperature independent and
consequently equal to the paramagnetic part of the zero-temperature energy, so
that  $\partial_{\rho_0} \F_\text{P} = \mu_\text{P}(\rho_0)$.

The ferromagnetic properties of the alloy (Eqs.(\ref{requirements})) determine
the Landau part of the free energy up to an unknown factor $f$:
\begin{eqnarray}
\Delta\F= f\left( 2 (\rho_0-\rho_\text{3d})^2 (\tilde T-1) (2s_0/\gamma)^2 +
(2s_0/\gamma)^4\right).\label{Landau}
\end{eqnarray}
The para-to-ferro shift of the chemical potential provided by $\Delta F$ is:
\begin{eqnarray*}
\Delta\mu(\rho_0,T) = -\partial_{\rho_0}\left(f(\rho_0-\rho_\text{3d})^4
(\tilde T-1)^2\right)\theta(1-\tilde T).
\end{eqnarray*}
The comparison of the previous equation at $T=0$ with
Eq.(\ref{ZeroTemperatureChemicalPotShift}), together with the assumption that
$f$ is also temperature independent, uniquely defines the function $f$:
\begin{eqnarray}
f(\rho_0) &=& (\rho_0-\rho_\text{3d})^{-2}/(4\Delta\nu).\label{ALandau}
\end{eqnarray}
\end{subequations}
Accordingly, $\Delta\mu(\rho_0,T)=\Delta\mu(\rho_0) (1-\tilde
T)\theta(1-\tilde T)$, with $\Delta\mu(\rho_0)$ from
Eq.(\ref{ZeroTemperatureChemicalPotShift}) (see Fig.\ref{Figure2}a).

Now, in the spirit of the orbital-free local spin density approximation (LSDA) we turn to an
inhomogeneous case by letting the electron and the spin densities spatially vary: $\rho_0\to\rho(\bm
x),s_0\to s(\bm x)$. Capacitively charging the conductor we make it an open system, which is governed by
the $\Omega$-functional:
\begin{subequations}
\label{OmegaPotential}
\begin{eqnarray}
\Omega = \int d^3\bm x\left({\cal W}+ \F_{\cal KX}- \mu_0 \rho + e \varphi
(\rho-\rho_\text{0})/2 \right).\label{OmegaPotential1}
\end{eqnarray}
Here $\F_{\cal KX}$ is defined by Eqs.(\ref{FreeEnergy}), $\mu_0 \equiv \mu(\rho_0,T_0)$, $\varphi=e\int
d^3\bm x'(\rho(\bm x')-\rho_0)/|\bm x -\bm x'|$ is the direct interaction potential with $e$ being the
electron charge, and the nonlocality of the kinetic energy functional is accounted for by the
$\alpha$-von Weiszacker term \cite{VonWeiszacker}, ${\cal W}= \alpha\hbar^2/(8m^*)\sum_\ud
(\bm\nabla\rho^\ud)^2/\rho^\ud_h$, with $m^*$ being the effective mass of the 3d holes and with
$\rho^\ud_h\equiv \rho_\text{3d}/2-\rho^\ud$ being the spin up and down hole densities. We adopt
$\alpha=1/9$ - the case when the $\alpha$-von Weiszacker approximation is asymptotically correct for
long wavelengths \cite{Succpt}, \emph{i.e.}, the domain of the applicability of the Ginzburg-Landau (GL)
theory below.

As we already mentioned, the TF layer's spatial electron density variation,
due to the injection of the carriers, is relatively small. Therefore, to be
able to reach the ferromagnetic region from an initially paramagnetic state,
we need an alloy which is close to its transition at $T_0$: $T_C\equiv
T_C(\rho_0)<T_0, \delta T = T_0 - T_C\ll T_0$ (see Fig.\ref{Figure2}b), so
that $\delta \rho = \rho_0 - \rho_C = \delta T/\kappa \ll \Delta \rho$, where
$\rho_C = \rho_\text{3d} - \Delta\rho$ and $\Delta\rho= T_0 /
\kappa\approx0.25/a^3$. Having chosen the composition, we can focus on a small
area on $(\rho,T)$ plane, such that $|\rho-\rho_C|\ll \Delta\rho$. Within the
area, $\rho_h^\ud\approx\Delta\rho/2$, and the von Weiszacker term simplifies
as
\begin{eqnarray}
{\cal W} \approx \alpha\hbar^2((\bm{\nabla} \rho)^2 + 4(\bm{\nabla}
s)^2)/(8m^*\Delta\rho).\label{vW}
\end{eqnarray}
\end{subequations}
In the Thomas-Fermi picture of a Fermi liquid, which the von Weiszacker correction relies on, the DOS at
the chemical potential is given as $\nu_\text{TF}=k_Fm^*/(2\pi^2\hbar^2)$, where
$k_F=(3\pi^2\Delta\rho)^{1/3}$ is the Fermi wave-vector. To make the von Weiszacker term energy-wise
consistent with the rest of the $\Omega$-potential, obtained by the physical arguments different from
the orbital-free LSDA, we can require that $\nu_\text{TF}=\nu_\ni$, or $m^*\approx 8.2 m_0$. This value
of the effective mass lies well within the wide range of the 3d sub-bands' effective masses, which vary
from several $m_0$ to almost $30 m_0$ (see, \emph{e.g.}, Ref.\cite{EffectiveNiMass}).

Even though the orbital-free LSDA has proven successful in some atomic-scale non-homogeneous problems,
in our case any atomic-scale results obtained from Eqs.(\ref{OmegaPotential}) would have a rather
qualitative character. Indeed, the very notion of the composition of a solid solution is well defined
only on scales larger than the lattice constant. Furthermore, the microscopic properties of the electron
subsystem of the TF-layer differ from those of the bulk and are interface-material dependent. The
$\Omega$-potential (\ref{OmegaPotential}), however, can be used to find the characteristic distances of
the spatial variations of $\rho$ and $s$, which are given by the the density and the spin correlation
lengths, $\lambda_\rho$ and $\lambda_s$. Omitting intermediate derivations, the correlation lengths
obtained from (\ref{OmegaPotential}) in a linear response manner are:
\begin{subequations}
\label{lambda_s}
\begin{eqnarray}
\lambda_\rho&=&2^{-1/2}\lambda_\text{TF}\text{Re}\sqrt{1+\sqrt{1-k_F/(\pi^2\Delta\rho\lambda_\text{TF}^2)}},\label{lambdap} \\
\lambda_s &=& \lambda_\rho(\Lambda/\delta T)^{1/2}.\label{lambdas}
\end{eqnarray}
\end{subequations}
Here the Thomas-Fermi radius $\lambda_\text{TF}=(4\pi e^2 (2\nu_\ni))^{-1/2}$
and $\Lambda = \alpha\kappa\gamma^2\Delta\nu/(4m^*\lambda_\rho^2)$.

The numerical estimations lead to $\Lambda\approx 7\times10^2$ K. In the
vicinity of the transition, the inequality $\delta T\ll \Lambda$ is well
satisfied, so that the spin and the density scales separate,
$\lambda_s\gg\lambda_\rho$ (see Eq.(\ref{lambdas})). Thus, we arrived at a
typical picture of the critical phenomena theory. The spin density is a
"ready-to-condense" \emph{soft} variable, behavior of which is governed by the
large-scale low-energy GL functional (the $s$-dependent part of
(\ref{OmegaPotential})), which in the $(T,\rho)$-area under consideration has
the following form:
\begin{eqnarray}
\F_\text{GL}=\int d^3\bm x\left(A(\bm \nabla s)^2 + b(\rho-\rho_C)s^2 + C
s^4/2\right),\label{GL}
\end{eqnarray}
with $b=2/(\gamma^4 \Delta\rho \Delta\nu)$, $A=k_s b$, $C=4b/(\gamma^2\Delta\rho)$. The high-energy
\emph{stiff} variable, $\rho$, plays a guiding role via the $\rho$-dependence of the effective chemical
potential for the magnetization (the overall coefficient in $s^2$-term of Eq.(\ref{GL}),
$B=b(\rho-\rho_C)$). The feedback action of the soft variable, $s$, on the stiff variable, $\rho$, is
weak and/or unimportant.

The proposed form of the near-critical $s$-$\rho$ coupling can be obtained on
a more general footing. Indeed, Taylor expanding $B$ around the transition
point, at which $B(T_0,\rho_0)=0$, and noticing that in itinerant ferromagnets
the $s$-$\rho$ coupling can only be local (especially on the
$\lambda_s$-scale), we arrive at $B\approx b(\rho-\rho_0)+b'(T_0-T_C)\equiv
b(\rho-\rho_C)$, with $\rho_C=\rho_0-\delta\rho, \delta\rho=(T_0-T_C)/\kappa,
\kappa \equiv b/b'$. For quantitative studies, however, it is crucial to
possess reliable values of the three material-specific parameters for
Eq.(\ref{GL}). The way the parameters for $\ni_\text{1-x}\cu_\text{x}$ are
derived in this Letter can now be summarized as follows: $A$ is obtained from
the orbital-free LSDA considerations (Eq.(\ref{vW})); the mutual relation
between $b$ and $C$ - from the empirical properties of the alloy and from the
Landau theory for the II-order phase transitions (Eq.(\ref{Landau})); the
overall energy factor for $b$ and $C$ - from previous ab-initio parameters of
Ni and Cu and from the Stoner theory of itinerant ferromagnetism
(Eqs.(\ref{ZeroTemperatureChemicalPotShift}) and (\ref{ALandau})).

On the $\lambda_s$-scale, the microscopic effects are scaled out. The
magnetization is ignorant to the microscopic details of the electron density
profile, \emph{e.g.}, the Friedel oscillations. The injected carriers' spatial
density in the TF layer, with a width of order of $\lambda_\rho$, can be
assumed infinitely narrow. Therefore, a positively biased surface can be
represented as a 1D semi-infinite solution ($z<0$, see Fig.\ref{Figure2}c) of
the GL Euler equation with $\rho(\bm{x})\to\rho_0$ and with the following
boundary condition imposed by the form of the $s$-$\rho$ coupling: $\lambda_s
\left.\partial_z \ln s \right|_{z=0} = \sigma/\sigma_\text{tr}$, where $\sigma
= - \int_{-\infty}^0 (\rho(z) - \rho_0)dz$ is the surface density of the
excess holes and $\sigma_\text{tr}=\lambda_s\delta\rho$ is the threshold
charging. The solution is $s = s^*(\sinh (\text{coth}^{-1}
(\sigma/\sigma_\text{tr}) - z/\lambda_s))^{-1} \theta(\sigma -
\sigma_\text{tr})$, where $s^* = ( 2 b \delta\rho / C )^{1/2}$. The magnetic
response of the surface can be characterized by the surface density of Borh
magnetons, $\m = \int_{-\infty}^0 2 s(z) d z$ (see Fig.\ref{Figure2}d): $\m=
\m^*\log
[\sigma/\sigma_\text{tr}+((\sigma/\sigma_\text{tr})^2-1)^{1/2}]\theta(\sigma-\sigma_\text{tr})$,
where $\m^* = (2A/C)^{1/2}$. The characteristic magnetization, $\m^*$, is independent of $\delta T$, so
that the effective magnetic susceptibility to the charging, $\m^*/\sigma_\text{tr}\propto\delta
T^{-1/2}$, grows infinitely as one approaches the transition point (see Fig.\ref{Figure2}d inset).

In reality, the surface charge is limited by the breakdown of the insulating interface material. For
SiO$_\text{2}$, the breakdown electric field $E_{\text{SiO}_2}\approx10^7$V/cm, which corresponds to the
following charging $\sigma_{\text{SiO}_2} = \epsilon E_{\text{SiO}_2} / (4 \pi e) \approx 2.5 \times
10^{13}\text{cm}^{-2}$, where $\epsilon=4.5$ is the SiO$_2$ dielectric constant. According to our
estimations, for the Ni$_\text{1-x}$Cu$_\text{x}$/SiO$_2$ interface, which is $\delta T=5$K away from
the transition, the threshold charging $\sigma_\text{tr}\approx0.15\times\sigma_{\text{SiO}_2}$, and on
the edge of the semiconductor breakdown ($\sigma\lesssim\sigma_{\text{SiO}_2}$) the surface
magnetization is $\m \approx 1.3\times10^{14}\mu_\text{B}\text{cm}^{-2}$.

The magnetic properties of the surface are in fact different from the bulk on their own, without
charging. This difference is modeled in the surface phase transition theory as a delta-functional jump
of the local Curie temperature at the surface of the semi-infinite system (see, \emph{e.g.},
Ref.\onlinecite{Diehl} and Refs. therein). Within the proposed approach the jump can be taken into
account as an intrinsic shift of the charging density $\sigma\to\sigma_\text{int}+\sigma$ (the cases
$\sigma_\text{int}\gtrless0$ are known respectively as ordinary and extraodinary transitions). A
reliable estimate for the material- and interface-dependent $\sigma_\text{int}$ can only be obtained
from ab-initio studies or experimental data.

Another issue is the quantum fluctuations, due to which the critical exponent of the spin correlation
length must acquire the renormalization group correction: $\lambda_s\to \lambda_s^* \approx
\lambda_\rho(T_C/\delta T)^{1/2+\delta}$ ($\delta\approx 0.14$ for, e.g., the $d=4-\epsilon$
approximation \cite{Ma}). Accordingly, the exponential tail of the magnetization profile will be
elongated and the magnetic response should acquire an enhancement factor $\propto(T_C/\delta T)^\delta$.
On the other hand, the intrinsic substitutional irregularity in a solid solution must shorten
$\lambda_s$ due to the Anderson localization mechanism. Near the transition, the quantum fluctuations'
effect is dominant and in reality the magnetic response of the surface is stronger than the one obtained
here on the mean-field level.

The authors are grateful to Lu J. Sham, Alan H. MacDonald, and Daniel
Neuhauser for useful comments and discussions. The work was supported in part
by Western Institute of Nanoelectronics at UCLA.


\begin{references}

\bibitem{SpintronicsReview}
S. A. Wolf \emph{et.al.}, Science {\bf 294}, 1488 (2001).
\bibitem{DMSGeneral} Y. D. Park \emph{et.al.},
Science {\bf295}, 651 (2002); I. \v{Z}uti\'{c} \emph{et.al.}, Rev. Mod. Phys. {\bf 76}, 323 (2004); A.
H. MacDonald, P. Schiffer, and N. Samarth, Nature Mat. {\bf 4}, 195 (2005); T. Jungwirth \emph{et.al.},
Rev. Mod. Phys. {\bf 78}, 809 (2006).
\bibitem{DMSDiscovery} H. Ohno \emph{et.al.}, Nature (London) {\bf408}, 944
(2000).
\bibitem{citeStoner} E. C. Stoner, Proc. Roy. Soc. {\bf A165},372 (1938).
E. P. Wohlfarth, Proc. Roy. Soc. {\bf A195}, 434 (1949).
\bibitem{NiCuAlloy} S. A. Ahern, M. J. Martin, and W. Sucksmith, Proc. R. Soc. (London) {\bf 248}, 145 (1958).
\bibitem{MinHam} L. Michalak, C. M. Canali, and V. G. Benza, Phys. Rev. Lett. {\bf 97}, 096804 (2006);
S. Kleff, J. von Delft, M. Deshmukh, and D.C. Ralph, Phys. Rev. B {\bf 64}, 220401(R) (2001); C. M.
Canali and A. H. MacDonald, Phys. Rev. Lett. {\bf 85}, 5623 (2000).
\bibitem{The_Book_On_Band_Structure}
D. A. Papaconstantopoulos, \emph{Handbook of the Band Structure of Elemental
Solids} (Plenum, NY, 1986). In the estimations, the position of the 3d band
edge, $\epsilon_\text{3d}$, is taken to be at the energy of the X5 $k$-point,
$\epsilon_\text{X5}$: $\mu_\text{P}(\rho_{\ni,\cu})-\epsilon_\text{3d}\approx
-0.21$eV and $1.57$eV respectively; the exchange splitting in Ni is:
$\epsilon_\text{X5}^\d-\epsilon_\text{X5}^\u\approx0.82$eV.
\bibitem{VonWeiszacker} R. G. Parr and W. Yang, \emph{Density-Functional Theory of Atoms and Molecules} (Oxford
University Press, New York, 1989).
\bibitem{Succpt} I. V. Ovchinnikov, L. A. Bartell, and D. Neuhauser, J. Chem. Phys. {\bf 126}, 134101 (2007).
\bibitem{EffectiveNiMass} S. Prakash and S. K. Joshi, Phys. Rev. B {\bf 2}, 915 (1970).
\bibitem{Diehl} H. W. Diehl, in \emph{Phase Transitions and Critical Phenomena}, vol 10, Eds. by C. Domb and J. L.
Lebowitz (Academic, London) (1986).
\bibitem{Ma} S.K. Ma, Rev. Mod. Phys. {\bf 45}, 589 (1973).
\end{references}
\end{document}